\documentclass[12pt]{iopart}

\catcode`\*11
\let\equation*\undefined
\let\endequation*\undefined
\catcode`\*12

\usepackage[utf8]{inputenc}
\usepackage{iopams,graphics,color,tikz}
\usetikzlibrary{datavisualization}

\newcommand\supp{\mathop{\mathrm{supp}}\nolimits}

\newcommand\erf{\mathop{\mathrm{erf}}\nolimits}
\renewcommand\Im{\mathop{\mathrm{Im}}\nolimits}

\begin{document}
\title{Vorticity and the birth of optical vortices}

\author{Václav Potoček$^1$, Yu Wang$^2$, Stephen M. Barnett$^2$}
\address{$^1$ Faculty of Nuclear Sciences and Physical Engineering, Czech Technical University in Prague, Břehová 7, 115 19, Praha 1, Czech Republic}
\address{$^2$ School of Physics and Astronomy, University of Glasgow, Glasgow G12 8QQ, United Kingdom}
\ead{vaclav.potocek@fjfi.cvut.cz}

\begin{abstract}
We exploit the vorticity, familiar from fluid mechanics and the theory of superfluids, as a tool to track the birth and subsequent development of optical vortices at a spiral phase plate.
\end{abstract}


\maketitle

\section{Introduction}

The quantitative study of vortices in fields has a long and interesting history.  In particular it was the work of Helmholtz \cite{Helmholtz} and Kelvin \cite{Kelvin,KelvinTait} in the nineteenth century that highlighted the significance of the vorticity in determining the properties of vortices in fluids and by the beginning of the twentieth century the study of vortices and vorticity were textbook material \cite{Lamb}.

The vorticity in fluid mechanics is simply the curl of the fluid velocity \cite{Batchelor,Landau,Saffman,Majda}.  It captures, in an essential way, the presence of a vortex as typically, it is zero away from a vortex but diverges on the axis of the core.  It is, therefore, a very powerful indicator of the presence and position of a vortex.  It is for this purpose that we use the vorticity in this paper, to investigate the birth of an optical vortex beam produced by a spiral phase plate \cite{Beijersbergen} to create a beam carrying orbital angular momentum \cite{Les,Book,Alison}

Vorticity has been applied to the study of vortices in fields other than classical fluid dynamics.  It has long been recognized that the vortices and vorticity play a crucial role in the properties of superfluids, where the superfluid velocity is associated with the gradient of the phase of the superfluid wavefunction \cite{Khalatnikov,Pines,Donnelly,Tilley,Vollhardt,Pismen,Pitaevskii,Leggett}.  The use of vorticity in optics has been pioneered by Berry and coworkers \cite{Berry98a,Berry98b,BerryDennis,Dennis} although the definition of vorticity used here differs slightly
from Berry's in taking a non-zero value only at a vortex core and, in this way, being closer to that employed in the theory of superfluids.  In optics there is no superfluid velocity, but there is an analogous relationship in scalar optics between the gradient of the complex field amplitude and Poynting's vector \cite{Born}.  The curl of this gradient is then the required vorticity.

In this paper we apply the vorticity to explore the creation of an optical vortex beam by an optical element.  Our particular choice is a spiral phase plate, in which the phase accumulated by light propagating through the plate increases linearly with the azimuthal angle.  Such plates necessarily have a discontinuity in thickness for once azimuthal angle and, as we shall see, this plays an important role in the generation of a vortex beam.  We note that ours is not the first study of this phenomenon.  In particular, Berry analysed the formation of vortices by such a phase plate \cite{Berry2004}, finding that the field amplitude at the position of the vortex generated is immediately suppressed on propagation.  This effect is associated with rapidly decaying evanescent components of the field.  In this paper we provide a complementary analysis to that of Berry by plotting the trajectory of the vortex from its birth at the phase plate.  To do so, we smooth the discontinuity in the thickness of the phase plate and study the associated evolution of vortices in the propagating field.
We find that in this regime vortices are born in pairs with opposite topological charges and that vortices of the ``wrong'' sign are ejected from the beam along the direction of the phase step.  We note that similar behaviour is seen in the propagation of fractional orbital angular momentum beams \cite{Berry2004,Joerg2007,Joerg2008}

\section{Optical vorticity}


In fluid mechanics the vorticity is simply the curl of the local fluid velocity \cite{Batchelor,Landau,Saffman,Majda}
\begin{equation}
\mbox{\boldmath$\omega$} = \mbox{\boldmath$\nabla$}
\times {\bf v} .
\end{equation}
The link between this and optical or quantum fields is most readily appreciated by appealing to the hydrodynamic formulation of quantum mechanics \cite{Bohm} in which a probability current can be defined by
\begin{equation}
{\bf j} = \frac{\hbar}{2m\rmi}\left(\psi^*\mbox{\boldmath$\nabla$}\psi - \psi\mbox{\boldmath$\nabla$}\psi^*\right),
\end{equation}
where $\psi$ is the wavefunction for the quantum particle of mass $m$.  There is a corresponding continuity equation for the probability density,
$|\psi|^2$, in the form
\begin{equation}
\frac{\partial}{\partial t}|\psi|^2 +
\mbox{\boldmath$\nabla$}\cdot{\bf j} = 0
\end{equation}
which, as pointed out by Bohm, is analogous to the local conservation of energy for electromagnetic waves as embodied in Poynting's theorem \cite{Jackson}:
\begin{equation}
\frac{\partial}{\partial t}w +
\mbox{\boldmath$\nabla$}\cdot{\bf S} = 0 ,
\end{equation}
where $w$ is the energy density and ${\bf S}$ is the Poynting vector.

It is important to realize that neither the probability current nor the Poynting vector can play a role analogous to the local velocity in fluid mechanics, if for no other reason than the incorrect dimensions of these quantities.  The link between the probability current and a local velocity was realized in the theory of superfluids simply by dividing the probability current by the local probability density, $|\psi|^2$ \cite{Khalatnikov,Pines,Donnelly,Tilley,Vollhardt,Pismen,Pitaevskii,Leggett,Landau41},
\begin{equation}
{\bf v} = \frac{\hbar}{2m|\psi|^2\rmi}\left(\psi^*\mbox{\boldmath$\nabla$}\psi - \psi\mbox{\boldmath$\nabla$}\psi^*\right)
= \frac{\hbar}{m}\mbox{\boldmath$\nabla$}
\arg\psi.
\end{equation}

In optics we can proceed by analogy with the treatment of superfluids.  In the eikonal approximation the wave is everywhere close to a local plane wave and, in this situation, we can write the Poynting vector in the form \cite{Born}
\begin{equation}
{\bf S} = c W \mbox{\boldmath$\nabla$}\arg u,
\end{equation}
where $W$ is again the local energy density and $u$ is the complex amplitude of the field treated as a scalar wave.  We arrive at a quantity analogous to the superfluid velocity simply by dividing by the energy density
\begin{equation}
{\bf v} = \mbox{\boldmath$\nabla$}\arg u 
\end{equation}
where, for simplicity, we also choose units in which the speed of light is unity $(c=1)$.

The vorticity for our optical field is then, by analogy with fluid mechanics, simply the curl of this quantity.  This definition requires some care, however, as we are well-accustomed to the idea that the curl of the gradient of a scalar field is zero.
For a stationary scalar optical field $u(\mathbf{r})$, we define vorticity as
\begin{equation}
\mbox{\boldmath$\omega$} =  \mbox{\boldmath$\nabla$}_w \times \mbox{\boldmath$\nabla$}_s \arg u.
\label{eq:v-def}
\end{equation}
Here $\mbox{\boldmath$\nabla$}_s$ 
is the nabla operator defined using strong derivatives, i.e., operating as a derivative on Lebesgue space of functions, 
while $\mbox{\boldmath$\nabla$}_w$ uses weak derivatives in the space of distributions \cite{Jost}. This distinction is important: if both derivatives were strong, or both weak, the curl of a divergence would be an exact zero in the 
corresponding space, although the intermediate steps would be different. Instead, we take the phase gradient as a strong derivative, which ignores any $2\pi$ jumps as well as phase singularities as they happen on sheets of measure zero. We convert the result, which for smooth functions $u$ is locally $L^1$-integrable, to a regular distribution. Finally, we take the curl of this distribution in $\mathcal{D}'$.

The result is always a singular vector field. It is defined on the support of $u$ and supported itself entirely on the vortex lines. In the neighbourhood of a vortex line passing $\mathbf{r}_0$ with tangent $\mathbf{n}$, where $u$ can be approximated to the least nonzero order in position, it locally behaves like
\begin{equation}
\mbox{\boldmath$\omega$}(\mathbf{r}) \equiv 2\pi s\,\delta^2\left(\mathbf{a}\cdot(\mathbf{r} - \mathbf{r}_0), \mathbf{b}\cdot(\mathbf{r} - \mathbf{r}_0)\right) \mathbf{n},
\label{eq:v-delta}
\end{equation}
where $\mathbf{a}$, $\mathbf{b}$ are two vectors forming an orthonormal triad with $\mathbf{n}$, $s \in \mathbb{Z}$ is the vortex strength and $\delta^2(\dots,\dots)$ denotes a two-dimensional delta function centred on the position where both its arguments are zero. This form of $\mbox{\boldmath$\omega$}$ can be verified by direct calculation.

Although the weak derivative may sound like a foreign or abstract concept, it should be noted that it has very convenient numerical properties. Indeed, the limit
\begin{equation}
\lim_{\varepsilon\to 0} \frac{f(x+\varepsilon)-f(x)}{\varepsilon}
\end{equation}
converges to the distributional derivative rather than to the strong one, for example, if $f(x)$ is a piecewise linear function, it approaches a sum of displaced delta distributions in the weak topology, rather than zero, which is the strong derivative of such $f$.

From the definition it immediately follows that
\begin{equation}
  \mbox{\boldmath$\nabla$}_w \cdot \mbox{\boldmath$\omega$} = 0,
\label{eq:v-div}
\end{equation}
which we interpret as an optical analogue of Helmholtz's second theorem: optical vortex lines don't have sources or sinks, they either form closed loops or extend to the boundary of $\supp u$, where they cease to be well defined.

A useful immediate result of the formulation \eref{eq:v-def} is the corresponding integral form,
\begin{equation}
\oint_{\delta A} \mbox{\boldmath$\nabla$}_s \arg u \cdot \rmd\mathbf{l} = \int_A \mbox{\boldmath$\omega$} \rmd\mathbf{S} = 2\pi \sum_{\mathbf r \in A} s_\mathbf{r},
\label{eq:v-int}
\end{equation}
where $s_\mathbf{r}$ are the strengths of the vortices passing through $A$, as in \eref{eq:v-delta}, and the positive sense of these is taken to be in the direction of the oriented area $A$.

\begin{figure}[t]
\centering
\includegraphics{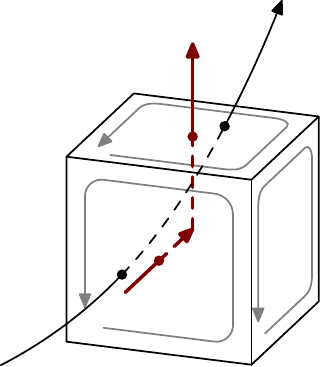}
\caption{The method to trace vortex lines numerically. We find integrals of phase gradients around faces of voxels and where these are nonzero we assume the normal vector to the face (red) as an approximation to the vorticity (black).}
\label{fig:cube}
\end{figure}

For practical calculations it is easier to use a discretized variant of the last equation than \eref{eq:v-def}. In this way we lose the access to the exact location and direction of the vorticity field, because we only choose a closed path and find whether (and in what strength, including sign) the field crosses the enclosed area. Nevertheless, we can choose small closed paths in planes perpendicular to coordinates and multiply the obtained integrals by corresponding unit vectors, placed at the paths' centroids, to get an approximant of the actual vorticity field. We shall use this method in the simulations below, where the paths are circling individual faces of cuboidal volume elements, so the approximant to the vorticity field is effectively a piecewise linear curve following the actual field within one distance unit (see \fref{fig:cube}).

Before turning to the evaluation of the vorticity for optical vortices, it is worth pausing to point out that the optical vorticity, as defined here, is not quite the same quantity as that introduced by Berry and Dennis \cite{BerryDennis}, which is
\begin{equation}
\mbox{\boldmath$\Omega$} = \frac{1}{2}\Im
\left(\mbox{\boldmath$\nabla$}u^*\times
\mbox{\boldmath$\nabla$}u\right) .
\end{equation}
The difference between the two quantities, {\boldmath$\omega$} and {\boldmath$\Omega$}, is 
readily appreciated by noting that both can be written as a curl, but of different quantities.  For 
{\boldmath$\Omega$} we have
\begin{equation}
\mbox{\boldmath$\Omega$} = \mbox{\boldmath$\nabla$}\times
\left(|u|^2\mbox{\boldmath$\nabla$}\arg u\right),
\end{equation}
which differs from the expression for {\boldmath$\omega$}, in \eref{eq:v-def}, by the presence of $|u|^2$.  This feature is responsible for very different properties displayed by the {\boldmath$\omega$} and {\boldmath$\Omega$}.  In particular, {\boldmath$\Omega$} can take a non-zero value anywhere and takes a finite value at the vortex core, but {\boldmath$\omega$} is zero everywhere except at a vortex core where it diverges.  Both quantities highlight the properties of a vortex, however: the component of {\boldmath$\Omega$} at a vortex core and in the direction of the vortex gives the vortex strength, but for {\boldmath$\omega$} it is the integral over a surface enclosing the vortex that reveals this quantity.
For our problem the vorticity {\boldmath$\omega$} is the more natural quantity to use because it is zero everywhere apart from at a vortex core and therefore provides a simple and effective way for finding and tracking vortices.  In this regard, moreover, it is the natural analogue of the vorticity as it is found in fluid mechanics and in the theory of superfluids.

\section{Analytical examples}

Consider a paraxial Laguerre-Gaussian beam propagating along the positive $z$ axis \cite{Les,Book,Alison,Siegman}:
\begin{equation}
\eqalign{
u_{\rm LG}(\rho, \phi, z) &= \left(\frac{2p!}{\pi(p+|\ell|)!}\right)^{1/2}\frac{(\rho\sqrt{2})^{|\ell|}}{w^{|\ell|+1}(z)}
\exp\left(-\frac{k_0\rho^2}{2(z_R + \rmi z)}\right) \\
& \qquad \times
L^{|\ell|}_p\left(\frac{2\rho^2}{w^2(z)}\right)\rme^{\rmi\ell\phi}
\exp\left(-\rmi(2p + |\ell| + 1)\tan^{-1}(z/z_R)\right),
}
\label{eq:lg}
\end{equation}
where we have used cylindrical coordinates, $w(z) = w_0/\sqrt{1 + z^2/z^2_R}$ and $z_R$ is the
Rayleigh range, $k_0w^2_0/2$.
In the following we will use a shorthand notation $\mathrm{LG}_{\ell,p}$ to refer to this function, and restrict ourselves for simplicity to $p = 0$.%
\footnote{For $p \ne 0$, the Laguerre polynomial $L_p^{|\ell|}$ will contribute to phase when it changes sign. However, sign changes do not contribute to the result, as discussed in more detail for Hermite-Gaussian beams below.}
The phase of this beam at position $\mathbf{r} = (x, y, z) = (\rho\cos\phi, \rho\sin\phi, z)$ has the form
\begin{equation}
\arg u(x, y, z) = \frac{k_0 \rho^2 z}{2(z^2 + z_R^2)} + \ell\phi + g(z),
\label{eq:LG-argu}
\end{equation}
where $g(z)$ encompasses terms dependent only on the $z$ coordinate. In Cartesian coordinates
the gradient of $\arg u$ is
\begin{equation}
\eqalign{
\mbox{\boldmath$\nabla$}_s \arg u =
&\left( \frac{k_0 x z}{z^2 + z_R^2} -\frac{\ell y}{x^2+y^2},
  \frac{k_0 y z}{z^2 + z_R^2} + \frac{\ell x}{x^2+y^2},
\right. \\ &\quad \left.
  g'(z) + \frac{\partial}{\partial z}\left( \frac{k_0 (x^2+y^2) z}{2(z^2 + z_R^2)} \right)
\right).}
\end{equation}
The regular part of the curl of this expression is of course zero, but the vorticity can be nonzero where the conventional derivatives diverge. This will not affect its $x$ and $y$ components, owing to the fact that the first term in \eref{eq:LG-argu} is entire, but $\omega_z$ is a distribution of the form
\begin{equation}
  \omega_z = \left( \frac{\partial}{\partial x} \right)_w \frac{\ell x}{x^2+y^2} + \left( \frac{\partial}{\partial y} \right)_w \frac{\ell y}{x^2+y^2},
\end{equation}
which acts on a test function $\varphi(x, y, z)$ as
\begin{equation}
  (\omega_z, \varphi) = -\left( \frac{\ell x}{x^2+y^2}, \frac{\partial\varphi}{\partial x} \right) - \left( \frac{\ell y}{x^2+y^2}, \frac{\partial\varphi}{\partial y} \right).
\end{equation}
On transforming to cylindrical coordinates, $\tilde{\varphi}(\rho, \phi, z) := \varphi(\rho\cos\phi, \rho\sin\phi, z)$ we find
\begin{equation}
\eqalign{
(\omega_z, \varphi) &= -\ell\int \cos\phi \left( \frac{\partial\tilde\varphi}{\partial\rho}\cos\phi - \frac{\partial\tilde\varphi}{\partial\phi} \frac{\sin\phi}{\rho} \right) \rmd\rho \rmd\phi \rmd z \\
&\qquad -\ell\int \sin\phi \left( \frac{\partial\tilde\varphi}{\partial\rho}\sin\phi + \frac{\partial\tilde\varphi}{\partial\phi} \frac{\cos\phi}{\rho} \right) \rmd\rho \rmd\phi \rmd z \\
&= -\ell\int \frac{\partial\tilde\varphi}{\partial\rho} \rmd\rho \rmd\phi \rmd z
= -\ell\int_\mathbb{R} \left( \int_{-\pi}^\pi \left[ \tilde\varphi(\rho,\phi,z) \right]_{\rho=0}^\infty \rmd\phi \right) \rmd z \\
&= 2\pi \ell \int \varphi(0, 0, z) \rmd z = 2\pi \ell\,(\delta^2(x,y), \varphi(x, y, z)).
}
\end{equation}
Therefore the vorticity is
\begin{equation}
  \mbox{\boldmath$\omega$}(x, y, z) = 2\pi \ell\,(0, 0, \delta^2(x,y)),
\end{equation}
as expected: $\mbox{\boldmath$\omega$}$ describes a singular vector field of strength $2\pi\ell$ times a delta peak in the direction of beam propagation (opposite if $\ell < 0$), that is a delta function 
positioned at the vortex core weighted by the phase accumulated on traversing a closed loop encircling the core.

As a second example we consider a Hermite-Gaussian beam \cite{Siegman}, for simplicity first-order and oriented such that the direction of propagation is $+z$ and the zero plane is $x = 0$: 
\begin{equation}
  u_{\rm HG}(x, y, z) = \sqrt{\frac{2}{\pi}}\frac{2x}{w^2(z)}
\exp\left(-\frac{k_0(x^2+y^2)}{2(z_R + \rmi z)}\right) \exp\left(-2\rmi\tan^{-1}(z/z_R)\right) .
\end{equation}
Here the phase can be written as
\begin{equation}
  \arg u(x, y, z) = \frac{k_0(x^2+y^2)z}{2(z^2 + z_R^2)} + \pi\theta(-x) + g(z),
\end{equation}
with considerations analogous to the above case and $\theta$ denoting the Heaviside step function. There is a phase jump of $\pi$ when crossing between positive and negative $x$ half-spaces, but for $x\ne0$,
\begin{equation}
\mbox{\boldmath$\nabla$} \arg u =
\left( \frac{k_0 x z}{z^2 + z_R^2},
  \frac{k_0 y z}{z^2 + z_R^2},
  g'(z) + \frac{\partial}{\partial z}\left( \frac{k_0 (x^2+y^2) z}{2(z^2 + z_R^2)} \right)
\right).
\end{equation}
As $x=0$ defines a zero-measure subset of $\mathbb{R}^3$, we can extend the validity of the last relation to all points in space if $\mbox{\boldmath$\nabla$}_s$ is used. Then it quickly follows that $\mbox{\boldmath$\omega$} = \mathbf{0}$ everywhere for this beam. In other words this Hermite-Gaussian mode, and indeed all the others, do not possess vorticity as we have defined it. This might seem surprising, given that Laguerre-Gaussian functions can be composed of Hermite-Gaussian ones, and vice versa. However, \eref{eq:v-def} is not linear in $u$, so there is nothing we can assume about a superposition given the vorticity of its individual terms.

\section{Birth of a vortex}

We have seen that the vorticity is divergenceless, \eref{eq:v-div}, and for this reason a vortex line in an almost-everywhere nonzero beam cannot start or end. In short, there are no sources or sinks of vorticity.  Given this, it is natural to ask, how does a vortex line come about or change its strength in an optical device manipulating the angular momentum, such as a spiral wave plate or a spatial light modulator \cite{Alison}? (We don't consider $q$-plates here \cite{Marrucci}, as their action cannot be described within scalar light theory.)

\begin{figure}[t]
\centering
\begin{tikzpicture}
\draw[very thin,color=black!20,step=0.784] (-3.2,-3.2) grid (3.2,3.2);
\draw[->] (-3.2,0) -- (3.2,0) node[right] {$\phi$};
\draw[->] (0,-3.2) -- (0,3.2) node[above] {$f_s(\phi)$};
\draw[thick,domain=-3.14:3.14,samples=1000,smooth] plot [id=f1] function{x - pi*erf(20*x)};
\draw[color=black!70,dashed] (-0.16,-3.2) -- (-0.16,3.2);
\draw[color=black!70,dashed] (0.16,-3.2) -- (0.16,3.2);
\node at (-3.14,0) [below] {$-\pi$};
\node at (-1.57,0) [below] {$-\pi/2$};
\node at (-.1,0) [below left] {$0$};
\node at (1.57,0) [below] {$\pi/2$};
\node at (3.14,0) [below] {$\pi$};
\node at (0.1,1.57) [right] {$\pi/2$};
\node at (0.1,3.14) [right] {$\pi$};
\node at (-0.1,-1.57) [left] {$-\pi/2$};
\node at (-0.1,-3.14) [left] {$-\pi$};
\draw[->,scale=.3] (1.4, 2) node[above right] {$\pi/s$} -- (0.6,0.05);
\draw[->,scale=.3] (-1.4, 2) node[above left] {$-\pi/s$} -- (-0.6,0.05);
\end{tikzpicture}
\caption{Graph of $f_{\ell,s}(\phi)$ for $s = 20, \ell = 1$.}
\label{fig:f1-graph}
\end{figure}
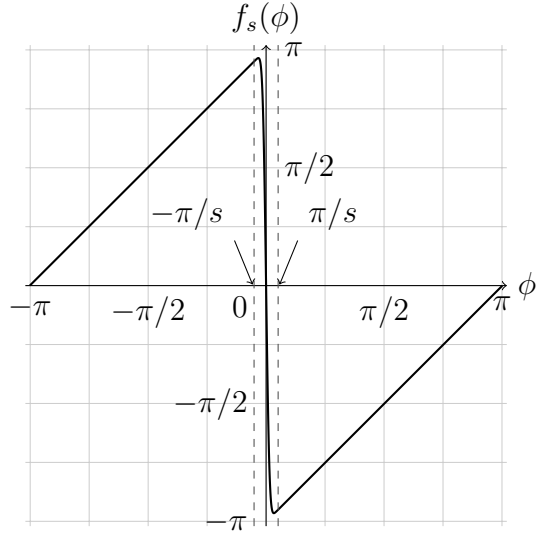

We address this question by considering the simplest explicit example; a paraxial light beam along $+z$ passing a hypothetical custom element modulating it at its output to the form
\begin{equation}
  u(\rho\cos\phi, \rho\sin\phi, 0) = \rho^{|\ell|} \rme^{\rmi f(\phi)} \rme^{-\rho^2/w_0^2},
\label{eq:initcond}
\end{equation}
where $f(\phi)$ is continuous over the unit circle but chosen such that in free-space evolution, this initial condition evolves to an approximation of a Laguerre-Gaussian beam. For the simplicity of our treatment, the radial dependence is already chosen according to the latter. An example of the angular dependence $f$ having the desired properties is
\begin{equation}
  f_{\ell,s}(\phi) = \ell(x - \pi\erf(s x))
\label{eq:fun}
\end{equation}
(see \fref{fig:f1-graph} for a graph).%
\footnote{This particular example is burdened by a slight discontinuity in phase from $f(\pi)$ to $f(-\pi)$, which decays super\-exponentially with $s$ and can safely be ignored in numerical calculation.}
Using such a function with a large value of $s$ in \eref{eq:initcond}, we see that the phase, following a circle around the beam axis, follows the same gradient as $\ell\phi$ for almost the full angle, but reverts to its starting value with a large negative gradient over a short interval. This way a global phase function can be defined and the initial condition is vortex-free. Nevertheless, in a cross section, the wedge in which the phase makes its return (interval around $\phi = 0$ in \fref{fig:f1-graph}) represents an area with large transversal $\mathbf{k}$ components. Thus it can be expected that this perturbation will eventually leave the bright part of the beam, leaving the dominating Laguerre-Gaussian background.

It should be noted that the higher value of $s$, the closer approximation \eref{eq:initcond} is to the Laguerre-Gaussian initial condition (in $L^2(\mathbb{R}^2)$) but at the same the higher transversal components of the $\mathbf{k}$ vector appear around $\phi = 0$. This is at odds with the assumptions of the paraxial approximation used, but can, for any fixed value $s$, be mitigated by choosing a larger beam waist size $w_0$. When then $w_0$ is used as a distance unit along $x$ and $y$ axes and the Rayleigh length $z_R$ as a distance unit along $z$ axis, the beam profile can in principle be scaled indefinitely while the solution expressed in these units converges to that described by the paraxial wave equation.

The expected behaviour of the initial condition \eref{eq:initcond} is confirmed by our computer simulation, performed using the numerical integration of the convolution thereof with the Fresnel kernel, an exact solution of the paraxial wave equation.

\begin{figure}[t]
\centering
\includegraphics[width=10cm]{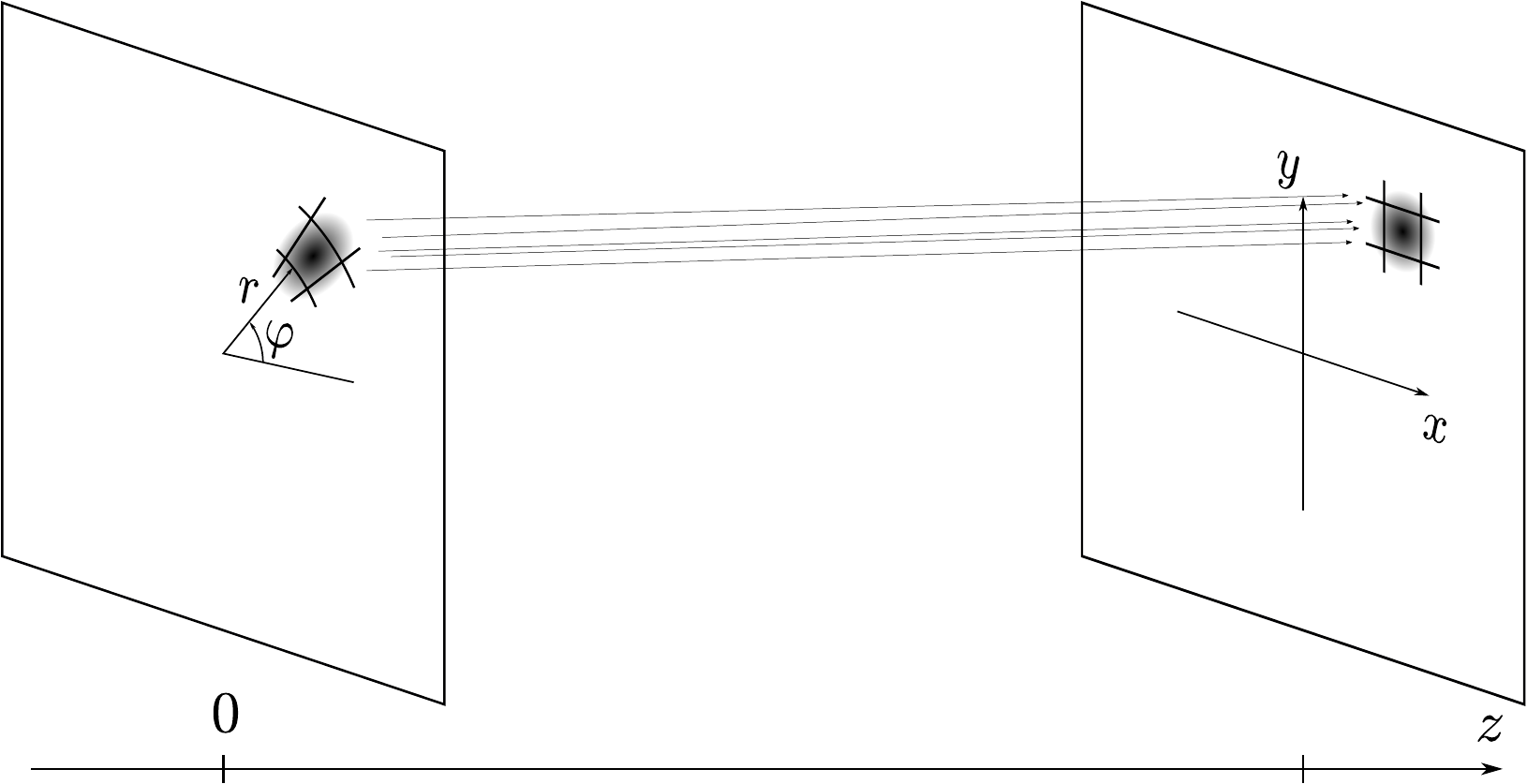}
\caption{Illustration of the numerical integration method. In the initial condition, $z=0$, polar coordinates are used, dividing the plane into wedge-shaped parts. In the target $z$ of the evolution, Cartesian coordinates are used for easier display and analysis. At both ends, a point source or a point detector is blurred using a Gaussian function, scaled and rotated to match the shape of the surface element.}
\label{fig:gauss}
\end{figure}

In order to represent the radial nature of the initial condition and avoid artificial effects of Cartesian position discretization, we perform the integration in polar source coordinates. The most interesting part of the evolution takes place very close to $z=0$ (in fractions of $z_R$), which prevents us from replacing the continuous source by a set of point sources, as diffraction artifacts obscure the result in such cases. We still sample the source function at a large number of polar lattice points, but use a small smooth radiating patch instead of a point source in each. For similar reasons, we convolve the calculated evolved wave front at $z = \mathrm{const}$ (which is sampled in an Euclidean lattice) with a mollifying function. For numerical convenience, we take Gaussian functions for both purposes (see \fref{fig:gauss}). As all the three operations -- point source blurring, evolution itself, and target plane smoothing -- are convolutions, they can simply be represented by adjusting the Fresnel convolution kernel accordingly. This simple modification removes most of the unphysical aliasing effects and allows our simulation to work down to one-ten-thousandth of $z_R$ splitting the radial and the polar coordinate in the initial condition in 500 and 1000 equal divisions, respectively.

In this method, each point of the target profile can be computed independently of others. In our simulation, we leverage this massive parallelizability by performing our calculations on the graphics processing unit, so although each 300 by 300 pixel output frame considers approximately $5\cdot10^{10}$ source--target coordinate pairs, it can be calculated in under one minute on a typical personal computer with integrated Intel$^{\rm(R)}$ graphics.

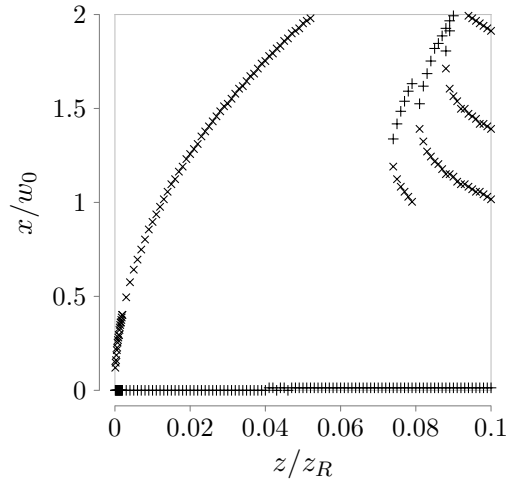
\begin{figure}[t]
\centering
\begin{tikzpicture}
\datavisualization [
scientific axes=clean,
all axes={ticks=few},
x axis={length=5cm, label={$z / z_R$}, min value=0, ticks={step=0.02}, /pgf/number format/fixed},
y axis={length=5cm, label={$x / w_0$}, max value=2, ticks=few},
visualize as scatter/.list={a,b,c},
style sheet=cross marks]
data [set=a,headline={x, y},read from file="data-l1-minus.table"]
data [set=b,headline={x, y},read from file="data-l1-plus.table"];
\end{tikzpicture}
\caption{The $x$ and $z$ coordinates of points with positive (plus sign) or negative (cross) $z$-components of vorticity for the $\ell=1$ case. Denser samples ($z_R/10^4$) are taken in the interval from $0$ to $0.002 z_R$, elsewhere $z_R/10^3$ is used. The points can easily be visually connected in vorticity lines, where the positive entries near $x=0$ and the negative ones form one line, breaking at right angle at the coordinate origin. Additional vortex--anti-vortex pairs appear further in the evolution and can similarly be attributed to intersections of curved vortex lines with the sampling plane. Lines which apparently end before the edge of the plotting region escaped the studied interval in the $y$ coordinate.}
\label{fig:l1-data}
\end{figure}

\begin{figure}[t]
\centering
\hbox to \columnwidth{%
\includegraphics[width=0.18\columnwidth]{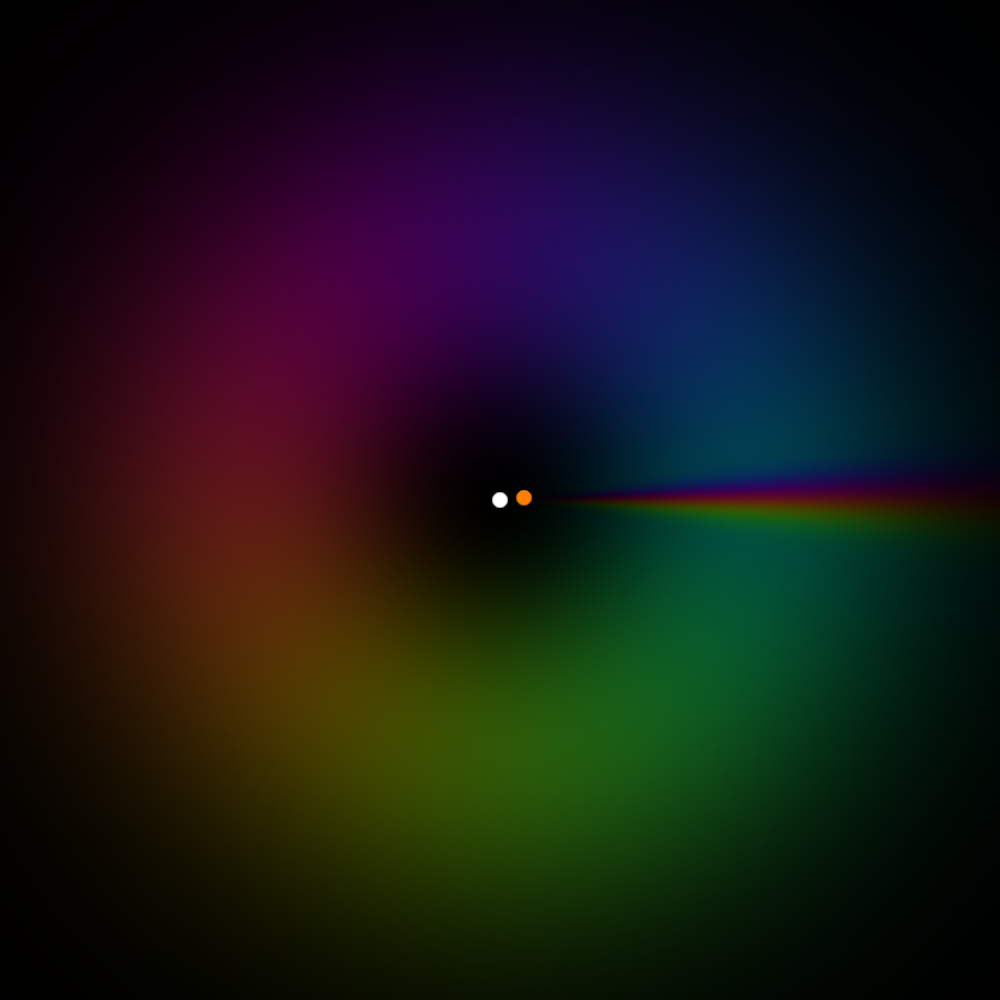}%
\hss
\includegraphics[width=0.18\columnwidth]{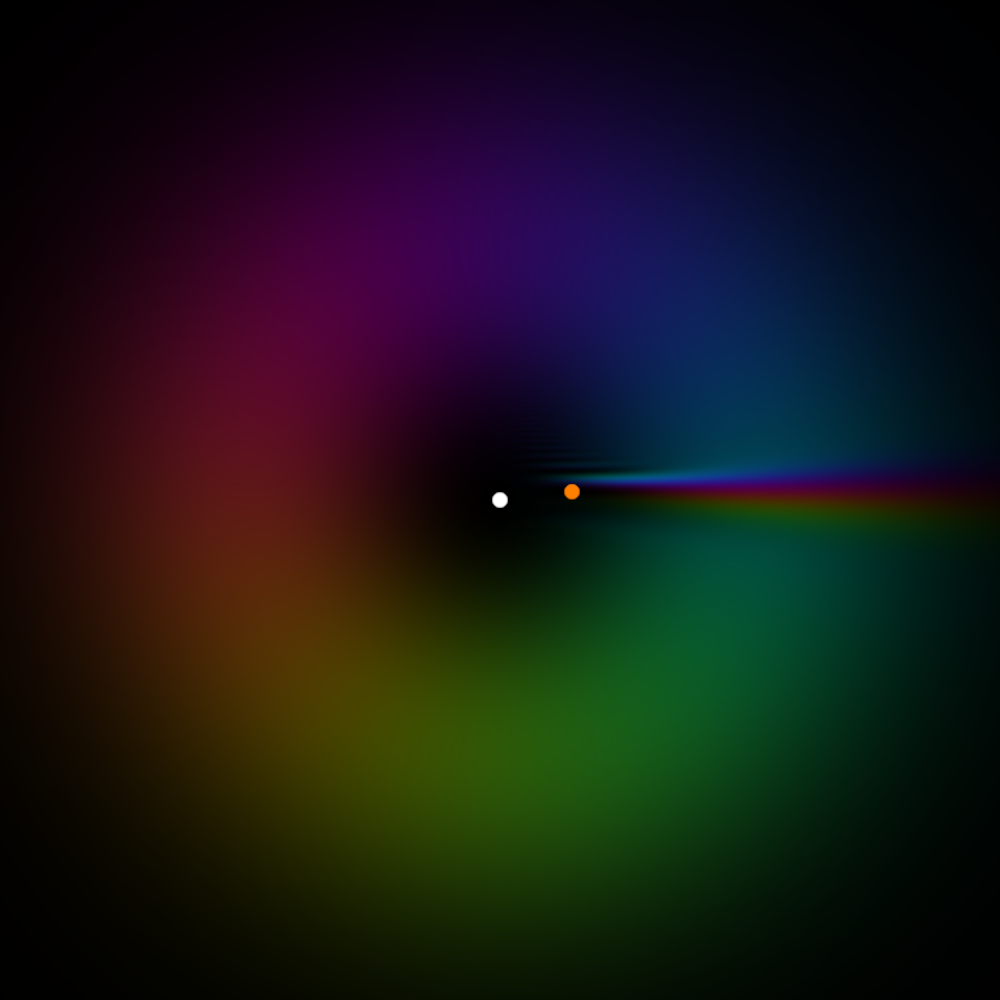}%
\hss
\includegraphics[width=0.18\columnwidth]{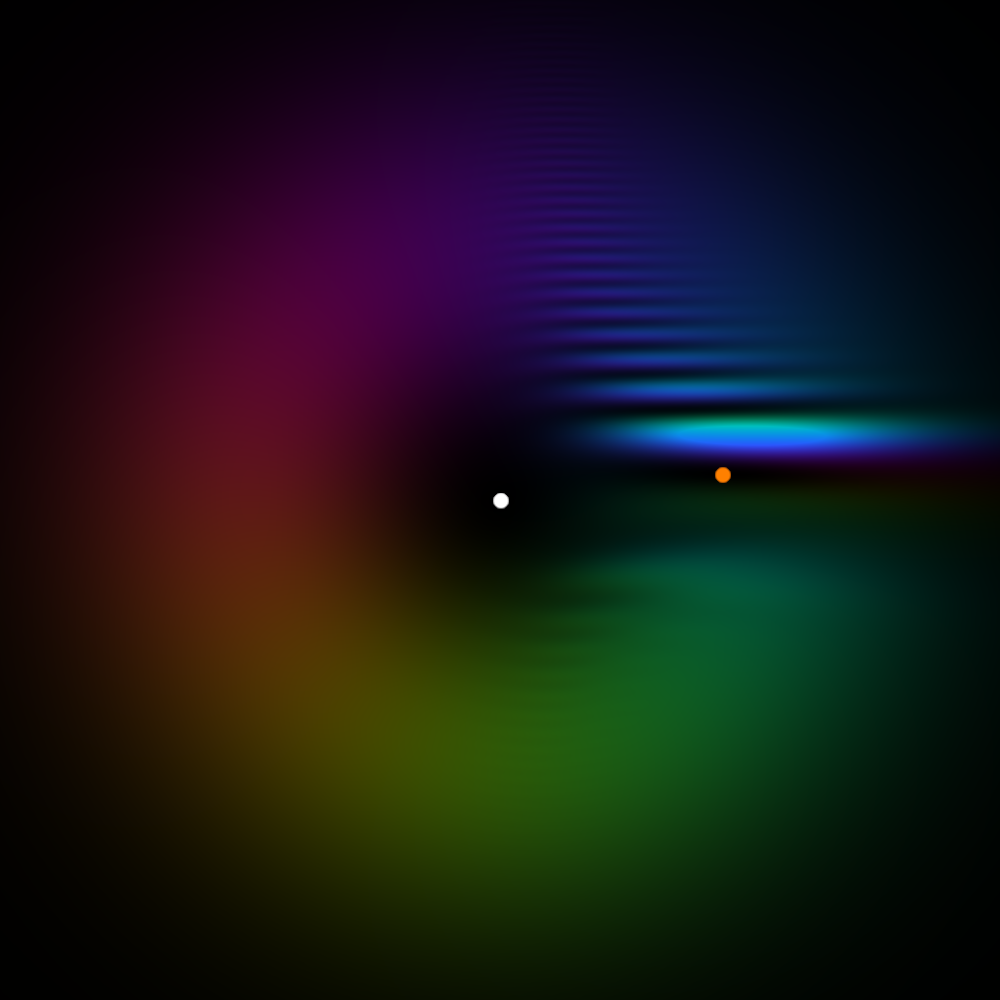}%
\hss
\includegraphics[width=0.18\columnwidth]{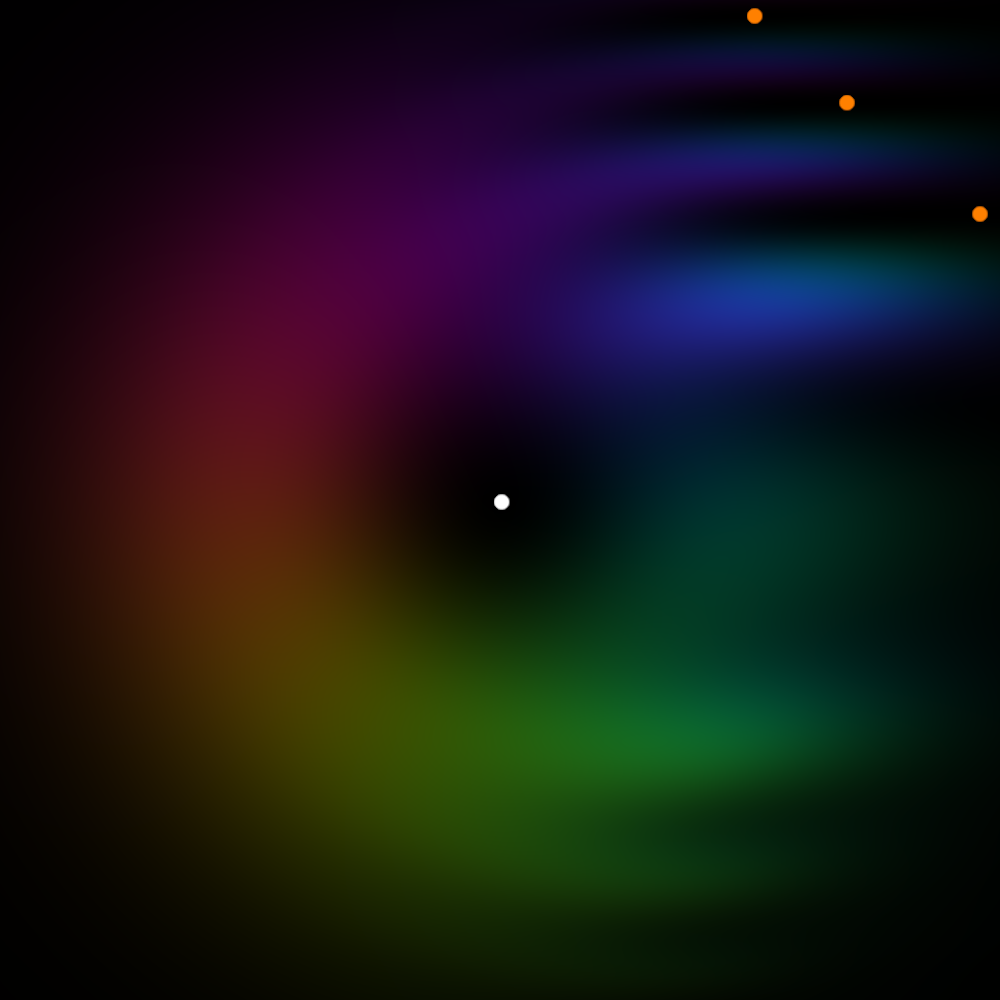}%
\hss
\includegraphics[width=0.18\columnwidth]{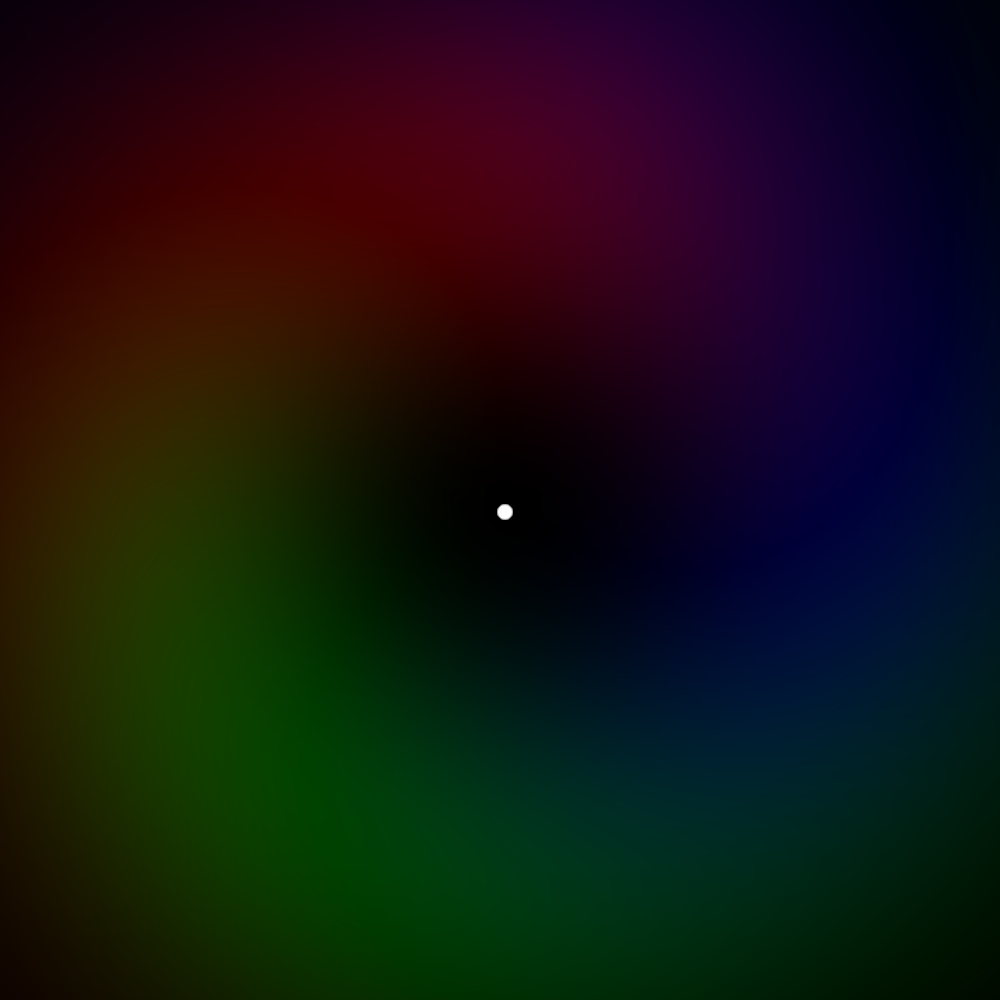}%
}%
\caption{Complex plots of $(x,y)$ slices of the field initiated by \eref{eq:initcond} with $\ell=1$, $f(\phi) = f_{1,10}(\phi)$ at different multiples of $z_R$: from left to right, $10^{-4}$, $10^{-3}$, $10^{-2}$, $10^{-1}$, $10^0$ (the latter adjusted in brightness). Hue and brightness represent the argument and squared magnitude of the value of $u(x, y, z)$, respectively. Distinctive white and red dots denote crossings with a vortex line with a positive and negative $z$ component, respectively.}
\label{fig:l1-frames}
\end{figure}

\Fref{fig:l1-data} shows our results for $\ell=1$, with $s=10$ in \eref{eq:fun}. We can observe a central vortex line, which is present at the earliest stages of the evolution reachable in our simulation. This line remains almost precisely in the origin and becomes the central zero of the formed Laguerre-Gaussian beam. It is accompanied, however, by another vortex line of opposite $\omega_z$. This rapidly escapes the bright portion of the beam: long before one tenth of Rayleigh length its displacement overcomes twice the beam width and continues growing. \Fref{fig:l1-frames} illustrates more concretely the ``anti-vortex'' forming near the main vortex and crossing the ring forming the beam: at $z$ very much less than $z_R$, the evolved function looks almost unchanged compared to the initial condition, but a pair of phase discontinuities is formed in the dark centre and these
rapidly separate, the distance between them tripling between $10^{-4}z_R$ and $10^{-3}z_R$.%
\footnote{It is important to keep in mind that various effects unaccounted for start to be important near the wavelength limit, so it would be unphysical to consider extremely small distances $z$. However, note that for $z_R = 1\,\mathrm{m}$, which is easily achievable, $10^{-4} z_R$ is still in hundreds of wavelengths for visible light.}
At $z = 10^{-2}z_R$, the disturbance to the phase profile distinctly forms a secondary wave that follows the direction of the phase gradient and interferes with the stable background. The anti-vortex finds its way through the ring in a local interference minimum, which is the only exact zero crossing in the otherwise bright region. At $z = 10^{-1}z_R$, the different transverse spatial frequencies in the perturbation separate, but all parts leave the main ring ballistically. The main anti-vortex has traversed well outside the studied interval. New vortex--anti-vortex pairs form in the outer parts of the ring temporarily. At $z = z_R$, an almost pure $\mathrm{LG}_{1,0}$ beam is obtained and all vortices except the central one leave the studied region and become part of the background noise. A broadening factor of $\sqrt{2}$, a radial deformation of the phase profile and an overall phase shift, as per \eref{eq:lg}, are clearly visible.

The main observation is that there is in fact a \textit{pair} of main vortex lines: the axis of the forming Laguerre-Gaussian beam with $\omega_z > 0$ and another line with $\omega_z < 0$ which rapidly passes out of the beam but can also be traced back almost to the origin of coordinates. Our simulation prevents us from following the line all the way to the origin -- there will always be a nonzero gap -- but we prove this below, along with the fact that the origin is reached perpendicular to $z$. We explain this observation that the vortex line near the optical axis did not start at $z=0$: it is but a part of a longer vortex line, that also came from $z = +\infty$ and turned, in a right angle, into the former.

\begin{figure}[t]
\centering
\begin{tikzpicture}
\datavisualization [
school book axes,
all axes={ticks=few},
x axis={min value=-1, max value=2, length=5cm, label={$x / w_0$}},
y axis={min value=-1, max value=1, length=3cm, label={$y / w_0$}},
visualize as scatter/.list={a,b},
style sheet=cross marks,
at start visualization={
\draw[color=red,dashed,<-] (visualization cs: x=0.2, y=0.0224) -- (visualization cs: x=1.8, y=0.202);
\draw[color=red,dashed,->] (visualization cs: x=0.2, y=-0.0224) -- (visualization cs: x=1.8, y=-0.202);
\node[red!50!black] at (visualization cs: x=1.64548, y=0.187291) [above] {$z=0.035z_R$};
\node[red!50!black] at (visualization cs: x=1.64548, y=-0.187291) [below] {$z=-0.035z_R$};
\node[red!50!black] at (visualization cs: x=0, y=0) [above left] {$z \to 0$};
}]
data [set=b] {
x, y
0.64214, 0.0668896
0.64214, -0.0668896
1.09699, 0.120401
1.09699, -0.120401
1.40468, 0.160535
1.40468, -0.160535
1.64548, 0.187291
1.64548, -0.187291
}
data [set=a] {
x, y
0, 0
};
\end{tikzpicture}
\caption{Positions of nonzero vorticity in $xy$ slices of the extended solution at several points near $z=0$, sampled in intervals of $0.01z_R$. The role of symbols was swapped compared to \fref{fig:l1-data} for easier readout. The mark at the origin is the central vortex line, present in all samples. We can see that the incidence angle of the anti-vortex undergoes an abrupt change at $z=0$.}
\label{fig:extension}
\end{figure}
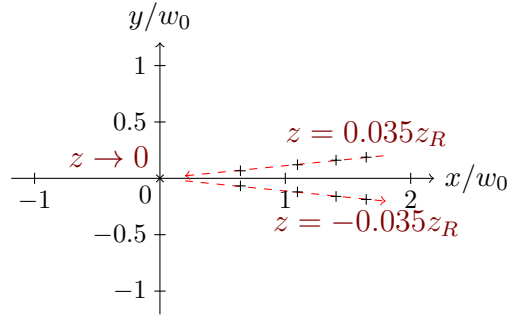

We note that if the free-space propagation was also extended to $z<0$, we could observe a central vortex line in that region as well. The anti-vortex would approach the origin as $z\to0_+$ and reappear in the negative-$z$ region (albeit at a different axial angle, see \fref{fig:extension}). It might seem natural to claim that two vortex lines of opposite $z$ components of vorticity crossed in the origin. However, it is entirely plausible that the initial condition \eref{eq:initcond} was formed from an incident vortex-less field (e.g., a Gaussian beam) in such way that the amplitudes and phases were smoothly modulated, and the field in $z<0$ indeed has no vortices. (It ceases to be an analytical continuation of our solution, because in $z>0$ we assume free space while $z<0$ is occupied by the material of the optical modulator.) In this case we are left with the interpretation of the anti-vortex line becoming the vortex line at the origin. There is a sudden right angle turn in this point, however, we note that, as observed above, there is a discontinuity in the direction of the vortex lines in either interpretation.

The graph, \fref{fig:l1-data}, suggests that the anti-vortex radial distance as a function of $z$ might be fitted by a multiple of the $\sqrt{z}$ function. We prove this observation analytically.

Let $E(z)$ denote the evolution operator of the paraxial wave equation for $z \in \mathbb{R}$,
\begin{equation}
  E(z) = \exp\left[ \rmi z \left( k_0 - \frac{1}{2k_0} \nabla_{x,y}^2 \right) \right],
\end{equation}
and further let $S(w)$ be a squeezing operator of the form
\begin{equation}
  S(w): \psi(x,y) \mapsto w \psi(w x, w y), \quad w>0.
\end{equation}
Both $E(z)$ and $S(w)$ are unitary on $L^2(\mathbb{R}^2)$ and satisfy
\begin{equation}
E(z) S(w) = S(w) E(z w^2).
\end{equation}
Now for $\rho\ll w_0$ the initial condition $\phi_0(x,y) := u(x,y,0)$ satisfies
\begin{equation}
  \phi_0(\rho\cos\phi, \rho\sin\phi) \approx \psi_0(\rho\cos\phi, \rho\sin\phi) := \rho^{|\ell|} \rme^{\rmi f(\phi)}
\end{equation}
where for $\psi_0$ it holds (as a functional identity, since $\psi_0 \not\in L^2(\mathbb{R}^2)$) that
\begin{equation}
  S(w) \psi_0 = w^{|\ell|+1} \psi_0,
\end{equation}
and thus
\begin{equation}
  S(w) E(zw^2) \phi_0 = E(z) S(w) \phi_0 \approx w^{|\ell|+1} E(z) \phi_0.
\end{equation}
Given that
\begin{equation}
  u(x, y, z) = \phi_z(x,y) := [E(z) \phi_0](x,y),
\end{equation}
we see that $u(x,y,z) = 0$ implies that $u(wx,wy,w^2z) \approx 0$. The error is small if $wx, wy$ remain much smaller than $w_0$, and thus the zeroes approximately follow the parametric curve
\begin{equation}
  (wx_0, wy_0, w^2z_0), \quad w>0
\label{eq:parabola}
\end{equation}
for some $(x_0, y_0, z_0)$, which describes a half-parabola.

\begin{figure}[t]
\centering
\hbox to \columnwidth{%
\includegraphics[width=0.18\columnwidth]{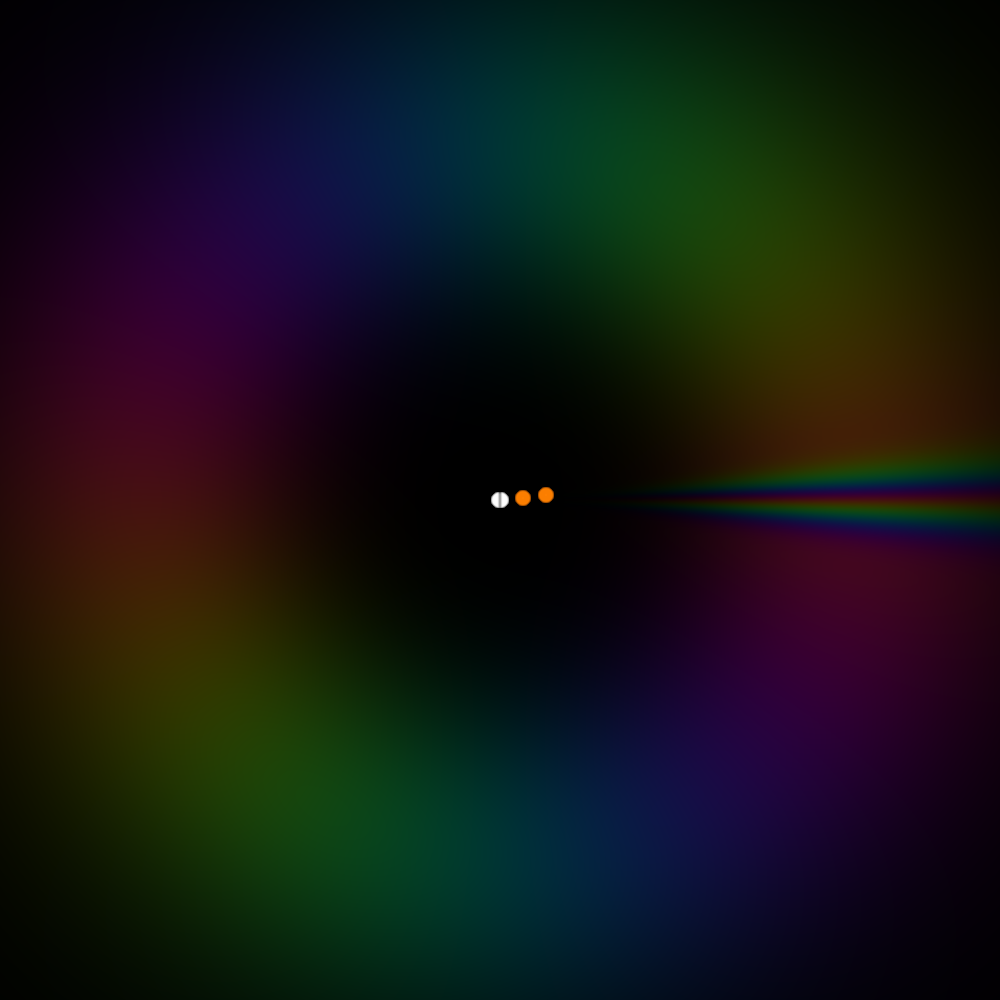}%
\hss
\includegraphics[width=0.18\columnwidth]{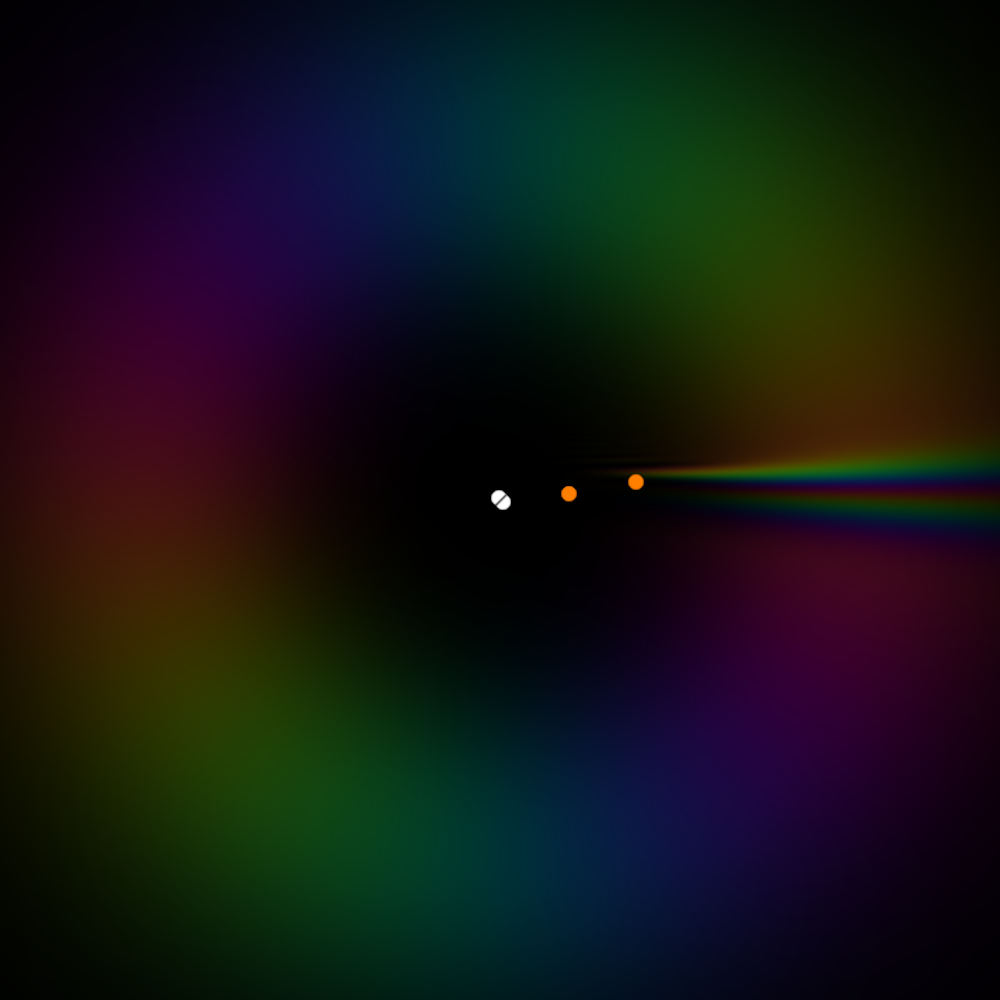}%
\hss
\includegraphics[width=0.18\columnwidth]{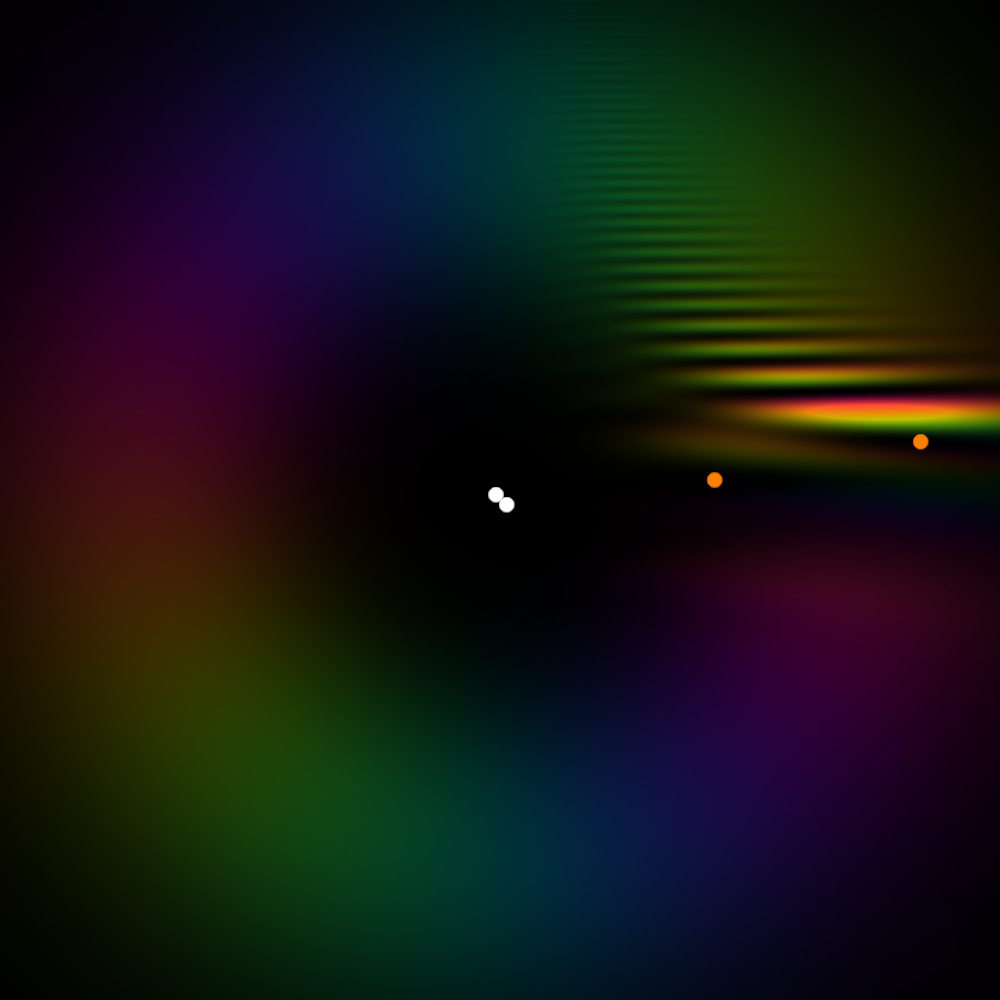}%
\hss
\includegraphics[width=0.18\columnwidth]{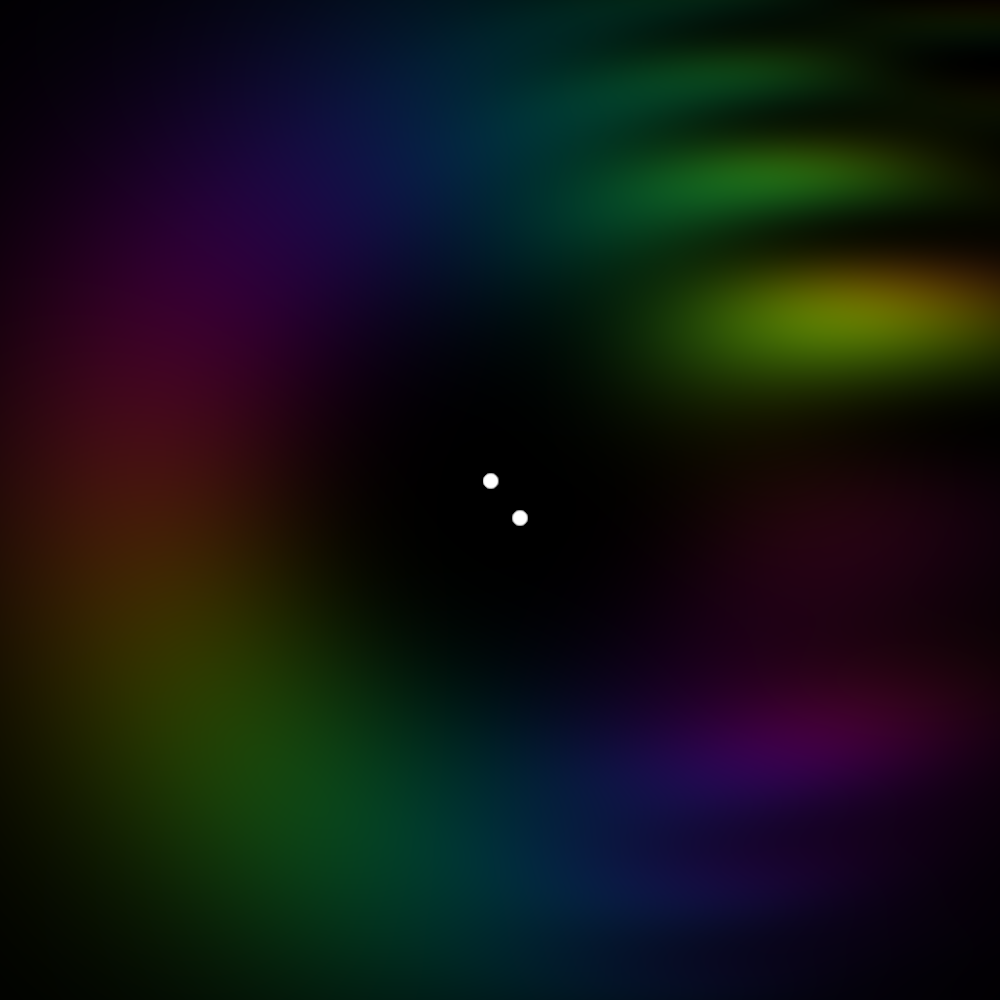}%
\hss
\includegraphics[width=0.18\columnwidth]{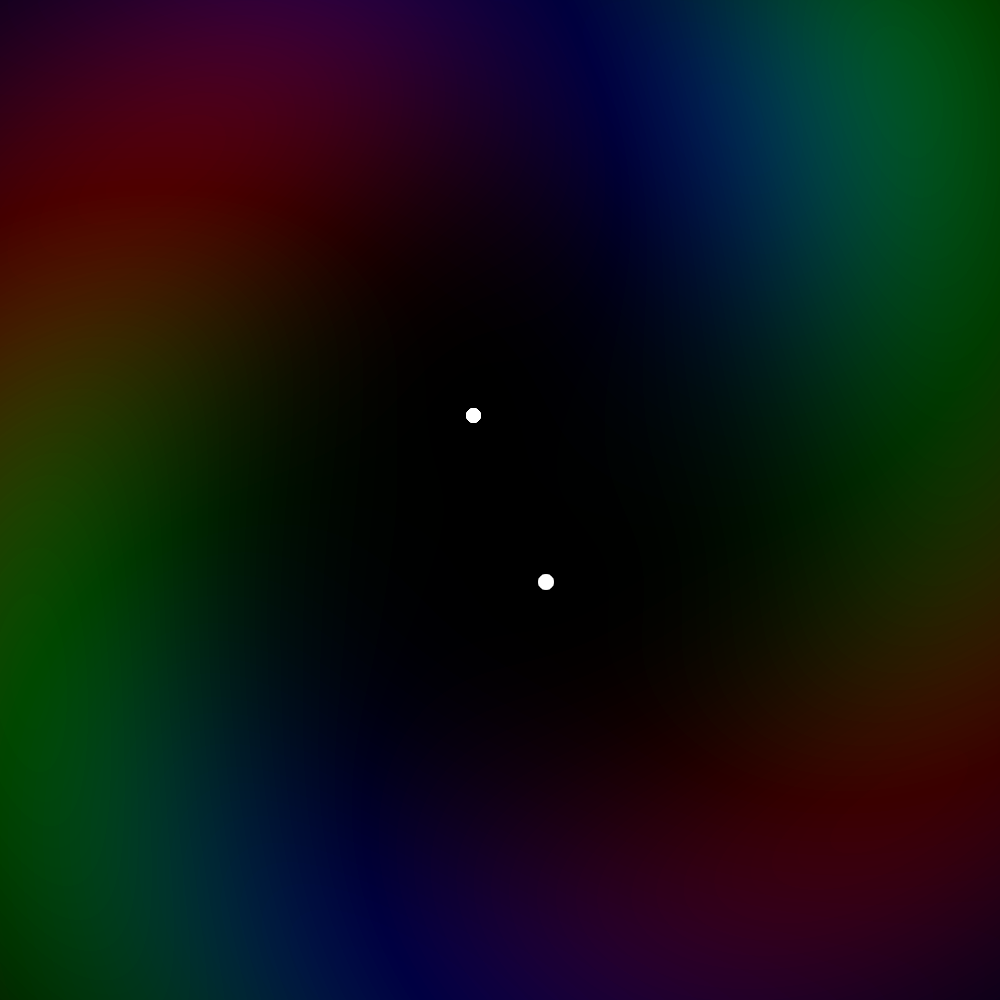}%
}%
\caption{Complex plots of slices of the field initiated by \eref{eq:initcond} with $\ell=2$, $f(\phi) = f_{2,10}(\phi)$ at different multiples of $z_R$: from left to right, $10^{-4}$, $10^{-3}$, $10^{-2}$, $10^{-1}$, $10^0$ (brightness adjusted). Color coding is the same as in \fref{fig:l1-frames}, a visual clue was added to better distinguish two nearby dots.}
\label{fig:l2u-frames}
\end{figure}

We also simulated the evolution of an initial condition corresponding to $\ell = 2$. Without the phase return of $f_{2,s}(\phi)$ (or in the limit in the mean as $s \to +\infty$), the initial condition \eref{eq:initcond} would describe a Laguerre-Gaussian beam $\mathrm{LG}_{2,0}$ and would have a vortex of double strength in the centre. Our main motivation was to see if this behaviour survives within the studied model, at least approximately. \Fref{fig:l2u-frames} shows that this is not the case: the central vortex is unstable under the perturbation and breaks into two single-strength vortex lines. As the reasoning behind \eref{eq:parabola} remains valid, each vortex line present near the origin will follow a parabolic trajectory, so we can conclude from the fact that two distinct near-central vortices are observed in the calculated samples of evolved wavefronts that the vortices lay on different parabolas, separated from the start.  The fact that our 
$\ell = 2$ beam has two $\ell = 1$ vortices rather than a single $\ell = 2$ vortex is a consequence of the well-known fact that higher-order optical vortices are unstable and, as has been shown experimentally \cite{vanExter}, realizations of higher $\ell$ vortex beams have, at their centres, $\ell$ charge $1$ vortices.

A total of two vortex and two anti-vortex lines meet in the origin of coordinates. In contrast with the case of $\ell = 1$, there seems to be no way to uniquely match the vortex and anti-vortex pairs: the only guarantee, given by \eref{eq:v-div}, is that the total incoming and outgoing vorticity sums up to zero in an arbitrarily small neighbourhood of each point. (On the other hand, any such pairing we might choose would be acceptable and consistent with the hypothesis that vortex lines do not start or end.)

An interesting side observation is that the anti-vortices leave at significantly different speeds -- a numerical fit suggests a ratio of 2 between their distances from the origin within the precision of the simulation -- and the same ratio can be observed between the distances of the two vortices from the origin. We do not yet have an explanation for this phenomenon. We also tried a modification of \eref{eq:fun} with $\ell=2$ where the phase return does not happen in one point but is split between $\phi=0$ and $\phi=\pi$: this resulted in a centrally symmetric arrangement of two vortices and two anti-vortices. Other than the disappearance of the difference in speeds the latter case did not provide any qualitatively different insights.

Supplementary material to this article contains animated evolutions of all the three simulated cases.

\section{Conclusions}

We have suggested a model of an imperfect spiral wave plate capable of generating Laguerre-Gaussian beams in the asymptotic regime from an initially vortex-less field. We showed that the central vortex, characteristic of the pure solution, does not appear alone, but rather is accompanied by a vortex of the opposite strength, which follows the disturbance and rapidly escapes the beam. In the three-dimensional space, both phase singularities can be understood to be parts of the same vortex line, which makes a sharp turn at the origin of coordinates.

This is consistent with a variant of the Helmholtz's second theorem, reformulated for scalar optics. However, as our second numerical example shows, it may not always be clear what parts constitute a given vortex line. The linearity of equation \eref{eq:v-div} allows superposition of different solutions, which may lead several vortex lines to meet precisely in a single point. In such cases one may choose some pairing between the incoming and outgoing trajectories to split them into pairs of independent vortex lines, or alternatively one may say that an arbitrary mixing took place in the intersection, only conserving the total topological charge.

The only places where a vortex line can end with no possible continuation within the medium is a boundary beyond which the wave function drops identically to zero. The definition of \eref{eq:v-def} tolerates sheets of zero intensity, as present for example in Hermite-Gaussian beams, but vorticity ceases to be well defined in open sets in which $u$ is zero. Boundaries of such regions then effectively act as an end of the medium. Nevertheless, should a region of zero intensity only appear inside an otherwise nonzero field (regardless on how such situation would be reached within the dynamics of an electromagnetic field), the integral form \eref{eq:v-int} guarantees that any vortex lines which may have struck it are recovered elsewhere on its boundary with the same total strength.

Finally, we note that the definition of $\mbox{\boldmath$\omega$}$ is not linked to any particular dynamical equation: we have used it to study solutions of the paraxial wave equation, but finding the solution and then analyzing it for vorticity are two independent steps. The quantity could be evaluated (and would retain its properties) with Helmholtz or the full wave equation, or even used to study a static wave function regardless of how it was obtained. Indeed, it is the same quantity as encountered in fluid dynamics and other parts of physics. Together with the ease of numerical calculation, this makes it a valuable tool in studying general scalar fields, without restriction on paraxiality, homogeneity, or properties of the initial conditions. The logical next step, though, would be to extend the definition to take into account polarization and vector singularities.

\ack

This research was funded in part by the EPSRC Grants No.\ EP/I012451/1, ``Challenges in Orbital Angular Momentum'' and No.\ EP/R008264/1 ``Relativistic Electron Vortices''. 
VP gratefully acknowledges funding by the Ministry of Education, Youth and Sports of the Czech Republic under grant number RVO 14000 and by the project ``Centre for Advanced Applied Sciences,'' Registry No.\ CZ.02.1.01/0.0/0.0/16\_019/0000778, supported by the Operational Programme Research, Development and Education, co-financed by the European Structural and Investment Funds and the state budget of the Czech Republic. SMB thanks the Royal Society for the award of a Research Professorship RP 150122.


\section*{References}

\begin{thebibliography}{99}%
\bibitem{Helmholtz} 
H. Helmholtz, \"{U}ber der hydrodynamische Gleichungen, welche den Wirbelbewegungen entsprechen, Journal f\"{u}r die reine und angewandte Mathematik {\bf 55}, 25-55 (1858).

\bibitem{Kelvin}
W. Thomson, On Vortex Motion, Trans. R. Soc. Edinburgh {\bf 25}, 217-260 (1869).

\bibitem{KelvinTait} 
W. Thomson and P. G. Tait, {\it Treatise on Natural Philosophy} vol. 1 (Cambridge University Press, Cambridge, 1883).

\bibitem{Lamb}
H. Lamb, {\it Hydrodynamics} 3rd ed. (Cambridge University Press, Cambridge, 1906).

\bibitem{Batchelor}
G. K. Batchelor, {\it An introduction to Fluid Mechanics} (Cambridge University Press, Cambridge, 1967).

\bibitem{Landau}
L. D. Landau and E. M. Lifshitz, {\it Fluid Mechanics} 2nd ed. (Elsevier, Amsterdam, 1987).

\bibitem{Saffman}
R. G. Safman, {\it Vortex Dynamics} (Cambridge University Press, Cambridge, 1992).

\bibitem{Majda}
A. J. Majda and A. L. Bertozzi, {\it Vorticity and Incompressible Flow} (Cambridge University Press, Cambridge, 2002).

\bibitem{Beijersbergen}
M. W. Beijersbergen, R. P. C. Coerwinkel, M. Kristensen and J. P. Woerdman, Helical-wavefront laser beams produced with a spiral phase plate, Opt. Commun. {\bf 112}, 321-327 (1994).

\bibitem{Les}
L. Allen, M. W. Beijersbergen, R. J. C. Spreeuw and J. P. Woerdman, Orbital angular momentum of light and the transformation of Laguerre-Gaussian laser modes, Phys. Rev. A {\bf 45}, 8185-8189 (1992).

\bibitem{Book}
L. Allen, S. M. Barnett and M. J. Padgett, {\it Optical Angular Momentum} (Institute of Physics Publishing, Bristol, 2003).

\bibitem{Alison}
A. M. Yao and M. J. Padgett, Optical angular momentum: origins, behavior and applications, Adv. Opt. Photon. {\bf 3}, 161-204 (2011).

\bibitem{Khalatnikov}
I. M. Khalatnikov, {\it An introduction to the theory of superfluidity} (Perseus Publishing, Cambridge MA, 2000).

\bibitem{Pines}
D. Pines and P. Nozi\'{e}res, {\it Theory of Quantum Liquids} (CRC press, Boca Raton, 2018).

\bibitem{Donnelly}
R. J. Donnelly, {\it Quantized Vortices in Helium II} (Cambridge University Press, Cambridge, 1991).

\bibitem{Tilley}
D. R. Tilley and J. Tilley, {\it Superfluidity and Superconductivity} 3rd ed. (Institute of Physics Publishing, Bristol, 1990).

\bibitem{Vollhardt}
D. Vollhardt and P. W\"{o}lfe, {\it The Superfluid Phases of Helium 3} (Taylor and Francis, London, 1990). 

\bibitem{Pismen}
L. M. Pismen, {\it Vortices in Nonlinear Fields} (Oxford University Press, Oxford, 1999).

\bibitem{Pitaevskii}
L. Pitaevski and S. Stringari, {\it Bose-Einstein Condensation} (Oxford University Press, Oxford, 2003).

\bibitem{Leggett}
A. J. Leggett, {\it Quantum Liquids} (Oxford University Press, Oxford, 2006).

\bibitem{Berry98a}
M. V. Berry, Wave dislocation reactions in non-paraxial Gaussian beams, J. Mod. Opt. {\bf 45}, 1845-1858 (1998).

\bibitem{Berry98b}
M. V. Berry, Much ado about nothing: optical dislocation lines (phase singularities, zeros, vortices ...), Proc. SPIE {\bf 3487}, 1-5 (1998).

\bibitem{BerryDennis}
M. V. Berry and M. R. Dennis, Phase singularities in isotropic random waves, Proc. R. Soc. Lond. A {\bf 456}, 2059-2079 (2000).

\bibitem{Dennis}
M. R. Dennis, K. O'Holleran and M. J. Padgett {\it Singular Optics: Optical Vortices and Polarization Singularities}, Progress in Optics {\bf 53}, 293-363 (2009).

\bibitem{Born}
M. Born and E. Wolf, {\it Principles of Optics} 6th ed. (Pergamon Press, Oxford, 1980).

\bibitem{Berry2004}
M. V. Berry, Optical vortices evolving from helicoidal integer and fractional integer phase steps, J. Opt. A: Pure Appl. Opt. {\bf 6}, 259-268 (2004).

\bibitem{Joerg2007}
J. B. G\"{o}tte, S. Franke-Arnold, R. Zambrini and S. M. Barnett, Quantum formulation of fractional orbital angular momentum, J. Mod. Opt. {\bf 54}, 1723-1738 (2007).

\bibitem{Joerg2008}
J. B. G\"{o}tte, K. O'Holleran, D. Preece, F. Flossmann, S. Franke-Arnold, S. M. Barnett and M. J. Padgett, Light beams with fractional angular momentum and their vortex structure, Opt. Expr. {\bf 16}, 993-1006 (2008).

\bibitem{Bohm}
D. Bohm, {\it Quantum Theory} (Dover, New York, 1989).

\bibitem{Jackson}
D. Jackson, {\it Classical Electrodynamics} (Wiley, New York, 1999).

\bibitem{Landau41}
L. Landau, Theory of the Superfluidity of Helium II, Phys. Rev. {\bf 60}, 356-358 (1941).

\bibitem{Jost}
J. Jost, \emph{Postmodern Analysis}. (Springer-Verlag, Berlin, 1997).

\bibitem{Siegman}
A. E. Siegman, \emph{Lasers} (University Science Books, Sausalito CA, 1986).

\bibitem{Marrucci}
L. Marrucci, C. Manzo, and D. Paparo, Optical Spin-to-Orbital Angular Momentum Conversion in Inhomogeneous Anisotropic Media,
Phys. Rev. Lett. {\bf 96}, 163905 (2006).

\bibitem{vanExter}
F. Ricci, W. L\"{o}ffler, and M.P. van Exter, Instability of higher-order optical vortices analyzed with a multi-pinhole interferometer, Opt. Expr. {\bf 20} 22961-22975 (2012).



\end{thebibliography}
\end{document}